\documentclass[10pt]{iopart}
\def\version{$$Id: plandet.tex,v 2.2 2003/03/06 10:45:32 karbach Exp $$}
\usepackage{graphicx,ifthen,iopams,setstack,amssymb,isolatin1}
\jl{1}
\newcommand{\writer}{michael}
\newcommand{\Kkappa}[2]{\mathcal{L}_{#1}(\{#2\})}

\newcommand{\KK}[2]{\mathcal{K}_{#1}(\{#2\})}

\eqnobysec
\begin{document}
%%%%%%%%%%%%%%%%%%%%%%%%%%%%%%%%%%%%%%%%%%%%%%%
\title{Transition rates via Bethe ansatz for the spin-1/2 planar XXZ antiferromagnet} 
\author{
  Daniel Biegel$^*$,
  Michael Karbach$^{*\dagger}$, and
  Gerhard M{\"u}ller$^\dagger$
}
\address{
  $^*$Bergische
  Universit{\"a}t Wuppertal, Fachbereich Physik,
  D-42097 Wuppertal, Germany \\
  $^\dagger$Department of Physics,
  University of Rhode Island,
  Kingston RI 02881-0817, USA
}
%\ead{michael@karbach.org, gmuller@uri.edu}
%\centerline{(\footnotesize \version)}

\ifthenelse{\equal{\writer}{gerhard}}%
{\date{\today~--~1.10}} % Gerhard
{\date{\version}}% Michael 
%\date{\today }
%%%%%%%%%%%%%%%%%%%%%%%%%%%%%%%%%%%%%%%%%%%%%%%
\begin{abstract}
  A novel determinantal representation for matrix elements of local spin
  operators between Bethe wave functions of the one-dimensional $s=\frac{1}{2}$
  $XXZ$ model is used to calculate transition rates for dynamic spin structure
  factors in the planar regime. In a first application, high-precision numerical
  data are presented for lineshapes and band edge singularities of the in-plane
  $(xx)$ two-spinon dynamic spin structure factor.
\end{abstract}
%\submitto{\JPA}
%\pacs{??}
%\maketitle
%%%%%%%%%%%%%%%%%%%%%%%%%%%%%%%%%%%%%%%%%%%%%%%
%
\section{Introduction}\label{sec:intro}
%
%%%%%%%%%%%%%%%%%%%%%%%%%%%%%%%%%%%%%%%%%%%%%%%
The importance of the spinon quasiparticle for the understanding of the quantum
fluctuations in integrable quantum spin chains is established on rigorous
grounds and further supported by experiments probing the dynamical properties of
quasi-one-dimensional magnetic compounds at low temperatures. Consider
the familiar and widely studied $s=\frac{1}{2}$ $XXZ$ antiferromagnet,
\begin{equation}\label{eq:HDelta}
  H\doteq J\sum_{n=1}^N \left(S_n^xS_{n+1}^x+S_n^yS_{n+1}^y+ \Delta S_n^zS_{n+1}^z \right),
\end{equation}
with $J>0, \Delta>0$. Its ground state for even $N$ can be configured as the physical
vacuum for spinons and its entire spectrum can be generated via systematic
creation of spinon pairs, for example, in the framework of the algebraic Bethe
ansatz \cite{FT84}.

Theoretical and experimental evidence points to the dominance of two-spinon
excitations in the low-temperature spin dynamics of the $XXZ$ antiferromagnet
\cite{IS80,MTPB81,GTN95,HSR+99,TCNT95}.
The exact two-spinon dynamic structure factor for the spin fluctuations
perpendicular to the antiferromagnetic long-range order in the axial regime
$\Delta>1$ was recently calculated via algebraic analysis \cite{KMB+97,BKM98}. This
method of exact analysis is not readily applicable in the planar regime $\Delta<1$,
but exact results for the case of isotropic exchange $(\Delta=1)$ can be inferred as
a limiting case. Extending the exact analysis of dynamic structure factors into
the planar regime has been a tantalizing challenge ever since.

Remarkable advances in the determination of matrix elements from Bethe wave
functions opened up new and promising avenues for the calculation of dynamical
properties of the $XXZ$ model. Of critical importance has been the recent work
of Kitanine, Maillet, and Terras \cite{KMT99} and the earlier work of Korepin
and Izergin \cite{Kore82,IK85a,KBI93}, which accomplished a reduction of the
norms of Bethe wave functions and of matrix elements (form factors) for the
local spin operators $S_n^\alpha$, $\alpha=x,y,z$, to determinants. We have already used
their formulas to calculate explicit expressions for transition rates of spin
fluctuation operators between eigenstates of the $XXX$ model $(\Delta=1)$ and have
applied them to calculate lineshapes of dynamic spin structure factors in a
magnetic field \cite{BKM02a}.  Here we present corresponding exact expressions
for the $XXZ$ model in the planar regime $(0<\Delta<1)$ with an application to the
in-plane two-spinon dynamic spin structure factor.  Related projects, which
focus on out-of-plane spin fluctuations, dimer fluctuations, $XX$ limit $(\Delta=0)$,
four-spinon structure factors, axial regime, and $T>0$ dynamics are in progress
\cite{note3}.

%%%%%%%%%%%%%%%%%%%%%%%%%%%%%%%%%%%%%%%%%%%%%%%
%
\section{Bethe ansatz equations}\label{sec:BAE}
%
%%%%%%%%%%%%%%%%%%%%%%%%%%%%%%%%%%%%%%%%%%%%%%%
The Bethe wave function of every eigenstate in the invariant subspace with
$z$-component $S_T^z=N/2-r$ of the total spin is specified by a set of
rapidities $z_1,\ldots,z_r$, which are solutions of the Bethe ansatz
equations \cite{CG66,note1}
\begin{eqnarray}\label{eq:BAEz}\fl
  N\arctan\left(\cot\frac{\gamma}{2}\tanh\frac{z_i}{2}\right) = \pi I_i
+\sum_{j\neq i}^r \arctan\left(\cot\gamma\,\tanh\frac{z_i-z_j}{2}\right),\;  i = 1,\ldots,r
\end{eqnarray}
with anisotropy parameter
\begin{equation}\label{eq:gamma}
\gamma\doteq \arccos\Delta~ ~ (0<\gamma<\pi).
\end{equation}

Different solutions within this subspace are distinguished by different sets of
Bethe quantum numbers $\{I_i\}$.  It is useful to further distinguish solutions
with only real magnon momenta $k_i$ and solutions where some or all $k_i$ are
complex.  The relation between the magnon momenta $k_i$ and the rapidities $z_i$
is
\begin{equation}\label{eq:ktoz}
\tanh\frac{z_i}{2} = y_i \doteq \tan\frac{\gamma}{2}\cot\frac{k_i}{2}.
\end{equation}

The transition rate expressions to be presented below are constructed to hold
for all Bethe ansatz solutions with real $k_i$. The generalizations necessary to
cover also solutions with complex $k_i$ are straightforward \cite{note4}.

It turns out that even in the restricted set of solutions with real $k_i$ some
of the rapidities may not be real, namely those with $|\tanh(z_i/2)|>1$, which
poses a problem in root-finding algorithms. To circumnavigate the problem we
introduce alternate rapidities $y_i$ defined in (\ref{eq:ktoz}).  They stay
real for all eigenstates with real $k_i$.  The Bethe ansatz equations for the
$y_i$ read
\begin{eqnarray}\label{eq:BAEy}\fl
N\arctan\left(y_i\cot\frac{\gamma}{2}\right) = \pi I_i 
\hspace*{-0cm} +\sum_{j\neq i}^r
\arctan\left(\cot\gamma\,\frac{y_i-y_j}{1-y_iy_j}\right),\;\;  i = 1,\ldots,r.
\end{eqnarray}
The wave number and energy of any given solution are
\begin{equation}\label{eq:wn}
k = \pi r -\frac{2\pi}{N}\sum_{i=1}^rI_i,
\end{equation}
\begin{eqnarray}\label{eq:enkzy}
\frac{E-E_F}{J} &=& -\sum_{i=1}^r(\Delta-\cos k_i) = -\sum_{i=1}^r\frac{\sin^2\gamma}{\cosh z_i-\cos\gamma},
\end{eqnarray}
where $E_F\equiv N\Delta/4$ is the energy of the reference state $|\!\!\uparrow\uparrow\cdots\uparrow\rangle$.

%%%%%%%%%%%%%%%%%%%%%%%%%%%%%%%%%%%%%%%%%%%%%%%
%
\section{Matrix elements}\label{sec:matel}
%
%%%%%%%%%%%%%%%%%%%%%%%%%%%%%%%%%%%%%%%%%%%%%%%
In generalization to the results presented in Ref.~\cite{BKM02a} for $\Delta=1$
we use the formulas for matrix elements $\langle\psi_0|S^{\mu}_n|\psi_\lambda\rangle$ from Kitanine,
Maillet, and Terras~\cite{KMT99} and the norms $\|\psi_\lambda\|$ from Korepin~\cite{Kore82}
of Bethe wave functions with real $k_i$ to calculate transition rates
\begin{equation}
  \label{eq:trarat}
  M_{\lambda}^{\mu}(q,\gamma) \doteq \frac{|\langle\psi_0|S^{\mu}_q|\psi_\lambda\rangle|^2}{\|\psi_0\|^{2}\|\psi_\lambda\|^{2}},\quad \mu=z,+,-
\end{equation} 
from the ground state of $H$ for the operators
\begin{equation}\label{eq:spiflu}
S_q^\mu = \frac{1}{\sqrt{N}}\sum_n\,e^{\rmi qn}S_n^\mu,\quad \mu=z,+,-.
\end{equation}
They probe the parallel $(\mu=z)$ and the perpendicular $(\mu=+,-)$ spin
fluctuations at zero temperature. An important and largely unanticipated feature
is that the determinantal expressions become much simpler in reciprocal space.
For the parallel spin fluctuations our calculations yield the following results:
\begin{eqnarray}
  \label{eq:21}\hspace*{-1.0cm}
  M_{\lambda}^{z}(q) &=&
  \frac{N}{4}\frac{\Kkappa{r}{z_{i}}}{\Kkappa{r}{z_{i}^{0}}}
   \KK{r}{z_{i}^{0}} \KK{r}{z_{i}}
   \frac{|\det(\mathsf{H}-\mathsf{P})|^{2}}%
        {\det \mathsf{K}(\{z_{i}\})\det\mathsf{K}(\{z_{i}^{0}\})},
\end{eqnarray}
where
\begin{equation}
  \label{eq:19}
  \Kkappa{r}{z_{i}} \doteq \prod_{i=1}^{r}\kappa(z_{i}), \quad
  \KK{r}{z_{i}} \doteq \prod_{i<j}^{r}|K(z_{i}-z_{j})|,
\end{equation}
\begin{equation}
  \label{eq:15}\hspace*{-1.5cm}
  \mathsf{H}_{ab} \doteq  \frac{\rmi}{2}\frac{\sin\gamma}{\sinh[(z^{0}_a - z_b)/2]} 
  \left(
     \prod_{j \neq a}^{r} G(z_j^0 - z_b) - d(z_b) \prod_{j \neq a}^{r} G^{*}(z_j^0 - z_b)
   \right), 
 \end{equation}
 \begin{equation}
\mathsf{P}_{ab} \doteq \rmi2\kappa(z_a^0) \prod_{j=1}^r G(z_j - z_b), \quad
a,b=1,\ldots,r,
 \end{equation}
\begin{equation}
  \label{eq:17}
  G(z) \doteq
  \sinh(z/2)\cot\gamma+\rmi\cosh(z/2),
  \end{equation}
  \begin{equation}
    d(z) \doteq \left(\frac{\tanh(z/2)\cot(\gamma/2)-\rmi}{\tanh(z/2)\cot(\gamma/2)+\rmi}\right)^{N},
  \end{equation}
\begin{equation} 
  \label{eq:12}
  \mathsf{K}_{ab} \doteq 
  \left\{
  \begin{array}[h]{ll}
    K(z_{a}-z_{b}) \cos\gamma      &: a\neq b \\ 
    N\kappa(z_{a})-\cos\gamma \sum_{j\neq a}^{r}K(z_{a}-z_{j}) &: a=b
  \end{array}\right.,
\end{equation}
\begin{equation}\label{eq:34}
  \kappa(z) \doteq \frac{1}{2}\frac{\sin^{2}\gamma}{\sinh^{2}(z/2)+\sin^{2}(\gamma/2)}, 
\end{equation}
\begin{equation}
  K(z) \doteq \frac{\sin^{2}\gamma}{\sinh^{2}(z/2)+\sin^{2}\gamma}.
\end{equation}
The corresponding results for the perpendicular spin fluctuations are
\begin{eqnarray}
  \label{eq:25}\hspace*{-2.0cm}
  M_{\lambda}^{\pm}(q) = 
  \left(\frac{\Kkappa{r}{z_{i}^{0}}}{\Kkappa{r\pm 1}{z_{i}}}\right)^{\pm1}
  \KK{r}{z_{i}^{0}}\KK{r\pm1}{z_{i}} 
  \frac{N|\det \mathsf{H^{\pm}}|^{2}}%
  {\det \mathsf{K}(\{z_{i}\})\det\mathsf{K}(\{z_{i}^{0}\})},
                                %\fl\nonumber \\ 
\end{eqnarray}
where
\begin{eqnarray} 
  \mathsf{H}^+_{ab} \doteq &\frac{\rmi}{2}\frac{\sin \gamma}{\sinh[(z_a - z^0_b)/2]} 
  \left(\prod_{j \neq a}^{r+1} G(z_j - z^0_b) - d(z^0_b) 
    \prod_{j \neq a}^{r+1} G^{*}(z_j - z^0_b) \right),
  \nonumber \\ 
  \mathsf{H}^+_{a,r+1} & \doteq \rmi \kappa(z_{a}), 
  \quad a=1,\ldots r+1, \; b=1,\ldots,r,
\end{eqnarray}
\begin{eqnarray} 
  \mathsf{H}^-_{ab} \doteq & \frac{\rmi}{2}\frac{\sin\gamma}{\sinh[(z_a^{0} - z_b)/2]} 
  \left(\prod_{j \neq a}^{r} G(z_j^{0} - z_b) - d(z_b) 
    \prod_{j \neq a}^{r} G^{*}(z_j^{0} - z_b) \right),
  \nonumber \\ 
  &&\nonumber\\ 
  \mathsf{H}^-_{ar} \doteq & \rmi \kappa(z_a^0), 
  \quad a=1,\ldots r, \; b=1,\ldots,r-1.
\end{eqnarray}
   
In the combined limits $\gamma\to 0, z_{i}\to 0, \; z_{i}/\gamma \to z_{i}'$, we recover
term by term the results reported in Ref.~\cite{BKM02a} for the $XXX$ model
$(\Delta=1)$. In the $XX$ limit $(\Delta=0)$ considerable simplifications occur in the
transition rate expressions but extreme care must be exercised in their
applications because of singularities in the Bethe ansatz equations. These
singularities manifest themselves, for example, in the occurrence of {\em
  critical} pairs of rapidities $y_i,y_j$ in Eqs.~(\ref{eq:BAEy}) with the
property $\lim_{\gamma\to0}y_iy_j=1$ \cite{Bieg00,DFM01,FM01}. Results for
the $XX$ model are forthcoming \cite{note3}.

More compact expressions for the results (\ref{eq:21}) and (\ref{eq:25}) can be
obtained as follows. Replacing (\ref{eq:12}) by the matrix
\begin{equation}
  \label{eq:bqn28}
  \bar\mathsf{K}_{ab} \doteq \left\{
  \begin{array}[h]{ll}
    {\displaystyle \frac{\cos \gamma}{N}\,\frac{K(z_{a}-z_{b})}{\kappa(z_{a})}}
    &: a\neq b \\  
    {\displaystyle 1-\frac{\cos \gamma}{N} \sum\limits_{j\neq a}^{r}\frac{K(z_{a}-z_{j})}{\kappa(z_{a})}} &: a=b
  \end{array}\right.
\end{equation}
allows us to extract a factor $N^r$ from $\det \mathsf{K}$:
\begin{equation}
  \label{eq:bqn29}
  \det \mathsf{K}(\{z_{i}\}) = N^{r} \Kkappa{r}{z_{i}} \det \bar\mathsf{K}(\{z_{i}\}).
\end{equation}
Using the Bethe ansatz equations in the algebraic form
\begin{equation}
  \label{eq:bqn11}
  d(z_{i}) =- \prod_{j=1}^{r} \frac{G^{*}(z_{i}-z_{j})}{G(z_{i}-z_{j})}
\end{equation}
again in $\mathsf{H}_{ab}$, $\mathsf{P}_{ab}$, and
$\mathsf{H}^\pm_{ab}$, further simplifies these matrices. The consolidated 
expressions then read
\begin{eqnarray}
  \label{eq:75}
  M_{\lambda}^{z}(q)  &= 
  \frac{N}{4} \frac{\KK{r}{z_{i}^{0}}}{\KK{r}{z_{i}}} 
  \frac{ |\det\left(\mathsf \Gamma - \frac{2}{N}\mathbf{1}\right)|^{2}}%
  {\det \bar \mathsf K(\{z_{i}^{0}\}) \det \bar \mathsf K(\{z_{i}\})},
  \\ \label{eq:76}
  M_{\lambda}^{\pm}(q) &=
  \left(
    \frac{\KK{r\pm 1}{z_{i}}}{\KK{r}{z_{i}^{0}}}
  \right)^{\pm1}
   \frac{|\det \mathsf{\Gamma^{\pm}}|^{2}}%
   {\det\bar \mathsf{K}(\{z_{i}\})\det \bar \mathsf{K}(\{z_{i}^{0}\})},
\end{eqnarray}
where
\begin{eqnarray}\fl
  \label{eq:77}
  \mathsf{\Gamma}_{ab} \doteq &
    F_{N}(z_{a}^{0},z_{b}) 
%    \Re 
    \left(
      \frac{1}{G(z_{a}^{0}-z_{b})}\prod_{j=1}^{r}\frac{G(z_{j}^{0}-z_{b})}{G(z_{j}-z_{b})}
      +
      \frac{1}{G^{*}(z_{a}^{0}-z_{b})}\prod_{j=1}^{r}\frac{G^{*}(z_{j}^{0}-z_{b})}{G^{*}(z_{j}-z_{b})}
    \right), 
    \\ \fl
  \mathsf{\Gamma}^+_{ab} \doteq& F_{N}(z_{a},z^0_b) 
 %\Re
  \left(
    \frac{G(z_{r+1}-z_{b}^{0})}{G(z_{a}-z^{0}_{b})}\prod_{j=1}^{r}\frac{G(z_j-z^0_b)}{G(z_j^{0} - z_b^{0})} 
    +
    \frac{G^{*}(z_{r+1}-z_{b}^{0})}{G^{*}(z_{a}-z^{0}_{b})}\prod_{j=1}^{r}\frac{G^{*}(z_j-z^0_b)}{G^{*}(z_j^{0} - z_b^{0})} 
   \right), \nonumber
   \\ \fl
  \mathsf{\Gamma}^+_{a,r+1} &\doteq  1, 
  \quad a=1,\ldots r+1, \; b=1,\ldots,r, 
  \\ \fl
  \mathsf{\Gamma}^-_{ab} \doteq & F_{N}(z_{a}^{0},z_b) 
  % \Re
  \left(
    \frac{G(z_{r}^{0}-z_{b})}{G(z_{a}^{0}-z_{b})}\prod_{j=1}^{r-1}\frac{G(z_j^{0} - z_b)}{G(z_j - z_b)} 
    +
    \frac{G^{*}(z_{r}^{0}-z_{b})}{G^{*}(z_{a}^{0}-z_{b})}\prod_{j=1}^{r-1} \frac{G^{*}(z_j^{0} - z_b)}{G^{*}(z_j - z_b)} 
  \right), \nonumber
  \\ \fl
  \mathsf{\Gamma}^-_{ar} \doteq & 1, 
  \quad a=1,\ldots r, \; b=1,\ldots,r-1,
\end{eqnarray}

\begin{equation}
  \label{eq:78}
  F_{N}(z,z') \doteq  \frac{\sin^{2}(\gamma/2)+\sinh^{2}(z/2)}{N\sin\gamma \sinh((z-z')/2)},  
\end{equation}
\begin{equation*}
  \label{eq:33}
  \mathbf{1} \equiv \left(
    \begin{array}[h]{ccc}
      1 & \cdots & 1 \\
      \vdots & \ddots  &\vdots  \\
      1 & \cdots  &1 \\
    \end{array}
  \right). 
\end{equation*}
In the isotropic limit $\gamma\to0$ we can use the results (\ref{eq:75})-(\ref{eq:77})
with 
\begin{equation}
  \label{eq:31}
  F_{N}(z,z') \doteq     \frac{1+z^{2}}{2N(z-z')},
\end{equation}
and the other ingredients as defined in \cite{BKM02a}. 

The corresponding results for $M_\lambda^z(q)$ and $M_\lambda^\pm(q)$ expressed in terms of
the rapidities $y_i$ are equally compact \cite{note3}. For applications
involving excitations with real magnon momenta $k_i$, we solve the Bethe ansatz
equations (\ref{eq:BAEy}) and calculate transition rates from (\ref{eq:75}) and
(\ref{eq:76}). 
   
%%%%%%%%%%%%%%%%%%%%%%%%%%%%%%%%%%%%%%%%%%%%%%%
%
\section{Two-spinon dynamic structure factor}\label{sec:2spdsf}
%
%%%%%%%%%%%%%%%%%%%%%%%%%%%%%%%%%%%%%%%%%%%%%%%
To demonstrate the practicality and usefulness of the newly derived transition
rate expressions, we present high-precision data for the
two-spinon part of the $T\!=\!0$ dynamic structure factor
\begin{equation}\label{eq:dssf}
S_{- +}(q,\omega) = 2\pi\sum_\lambda M_\lambda^-(q)\delta\left(\omega-\omega_\lambda\right),
\end{equation}
which captures the in-plane spin fluctuations. The state $|\psi_0\rangle$ in
(\ref{eq:trarat}) is the $XXZ$ ground state (spinon vacuum). It has quantum
number $S_T^z=0$ and is characterized by the solutions of Eqs.~(\ref{eq:BAEy})
for the set of $r=N/2$ Bethe quantum numbers $I_{i}^{(0)}=-(N+2)/4+i,
i=1,\ldots,N/2$.  Excited states containing two spinons with parallel spins are known
to account for a major contribution to the in-plane spin fluctuations
\cite{MTPB81}.

The number of two-spinon states with $S_T^z=1$ is $\frac{1}{8}N(N+2)$. Their
$r=N/2-1$ Bethe quantum numbers, which are integers for odd $r$ and
half-integers for even $r$, comprise all configurations
\begin{equation}\label{eq:2spt1}
-\frac{N}{4} \leq I_1 < I_2 < \cdots < I_r \leq \frac{N}{4}. 
\end{equation}
These states are described by real solutions of Eqs.~(\ref{eq:BAEy}). For $N\to\infty$
they form a continuum in $(q,\omega)$-space with boundaries \cite{CG66,MTPB81}
\begin{equation}\label{eq:2spelu}
\epsilon_L(q)=\frac{\pi J\sin\gamma}{2\gamma}\left|\sin q\right|,\qquad \epsilon_U(q)=\frac{\pi
  J\sin\gamma}{\gamma}\left|\sin \frac{q}{2}\right|. 
\end{equation}
The scaled density of two-spinon states constructed from $D^{(2)}(q,\omega_\lambda)\equiv 
2\pi/[N(\omega_{\lambda+1}-\omega_\lambda)]$ for the energy-sorted sequence of two-spinon excitations at
fixed $q$ and finite $N$ turns into the function
\begin{equation}\label{eq:2spdof}
D^{(2)}(q,\omega)=[\epsilon_U^2(q)-\omega^2]^{-1/2}
\end{equation}
for $\epsilon_L(q)\leq \omega\leq \epsilon_U(q)$ in the limit $N\to\infty$. For the two-spinon part of the
dynamic spin structure factor we use the product representation
\begin{equation}\label{eq:prodanpm}
S_{- +}^{(2)}(q,\omega)=M_{- +}^{(2)}(q,\omega)D^{(2)}(q,\omega)
\end{equation}
as discussed in a previous application of a similar nature \cite{KM00}. The
first factor represents the scaled transition rates  $M_{- +}^{(2)}(q,\omega)=N|\langle
G|S_q^-|\lambda\rangle|^2$ between the ground state and the two-spinon states
(\ref{eq:2spt1}).

\begin{figure}[htbp]
  \centering%\hspace*{40mm}
  \includegraphics[width=70mm,angle=-90]{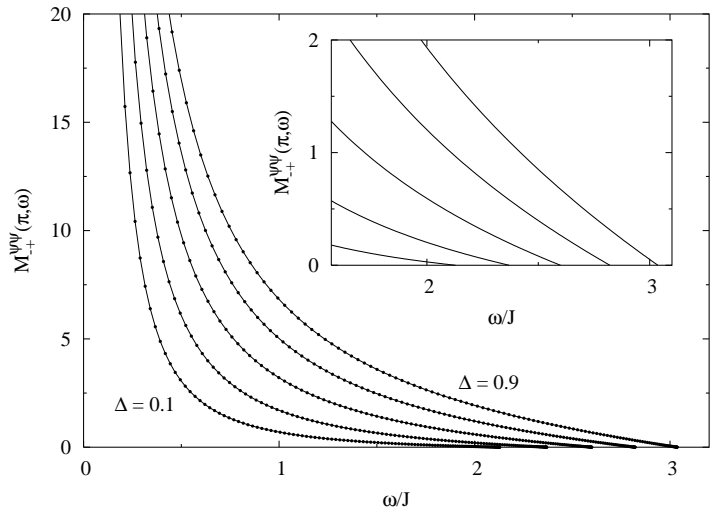}  
%\vspace*{30mm}
  \caption{Scaled transition rates between the $XXZ$ ground state and the
    two-spinon states at $q=\pi$ for $N=512$ and $\Delta=0.1, 0.3, \ldots, 0.9$. The inset
    shows the same data again for $\omega$ near $\epsilon_U(\pi)$ on a different scale.}
  \label{fig:fig1}
\end{figure}

Here we focus on the lineshape and on the singularity structure of $S_{-
  +}^{(2)}(q,\omega)$ at $q=\pi$, where the spectral threshold is at $\omega=0$. In
Fig.~\ref{fig:fig1} we show $N=512$ data for the transition rate function $M_{-
  +}^{(2)}(\pi,\omega)$ at $\Delta=0.1, 0.3, \ldots, 0.9$. The data indicate an infrared
divergence and exhibit a monotonically decreasing $\omega$-dependence toward zero
intensity at the upper band edge. The inset to Fig.~\ref{fig:fig1} zooms into
the behavior near $\epsilon_U(\pi)$. The data points approach zero linearly with a slope
that becomes smaller with decreasing $\Delta$. A linear behavior was already
established via algebraic analysis for the case $\Delta=1$ \cite{KMB+97}.
%%%%%%%%%%%%%%%%%%%%%%%%%%%%%%%%%%%%%%%%%%%%
%%%%%%%%%%%%%%%%%%%%%%%%%%%%%%%%%%%%%%%%%%%%%%%

\begin{figure}[htb]
  \centering
  \includegraphics[width=70mm,angle=-90]{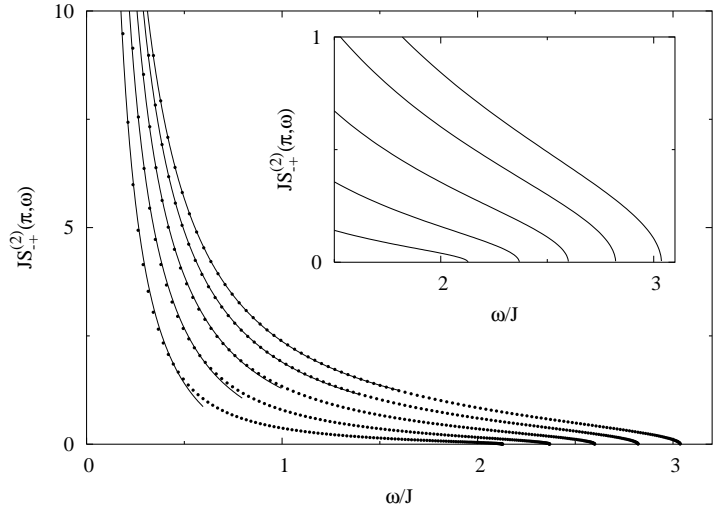}  
  \caption{Lineshape at $q=\pi$ of the two-spinon contribution to $S_{-
      +}(q,\omega)$. The dots represent data for $N=512$ at $\Delta=0.1, 0.3, \ldots, 0.9$
    (left to right). The lines are two-parameter fits $a\omega^{-1-\gamma/ \pi}+b$ (over the
    interval shown) with the exactly known exponent from Eq.~(\ref{eq:smpexp})
    for $N=\infty$. The inset shows the $N=512$ data again for $\omega$ near $\epsilon_U(\pi)$ on a
    different scale.}
  \label{fig:fig2}
\end{figure}

The spectral-weight distribution of the two-spinon contribution to $S_{-
  +}(\pi,\omega)$ is then inferred via product representation (\ref{eq:prodanpm}) from the
transition rate data and the two-spinon density of states (\ref{eq:2spdof}). The
results are shown in Fig.~\ref{fig:fig2}. The divergent density of states
(\ref{eq:2spdof}) at $\omega=\epsilon_U(\pi)$ converts the linear cusp of the function $M_{-
  +}^{(2)}(\pi,\omega)$ into a square-root cusp in the function $S_{- +}^{(2)}(\pi,\omega)$ as
is best visible in the inset.  The $\Delta$-dependent exponent of the power-law singularity is
exactly known \cite{LP75b,FGM+96}:
\begin{equation}\label{eq:smpexp}
S_{- +}(\pi,\omega) \sim \omega^{-1-\gamma/ \pi}.
\end{equation}
This non-universal critical singularity is accurately reflected by the $N=512$
data as is demonstrated by a data fit at low frequencies for each lineshape.
%%%%%%%%%%%%%%%%%%%%%%%%%%%%%%%%%%%%%%%%%%%%%%%
%%%%%%%%%%%%%%%%%%%%%%%%%%%%%%%%%%%%%%%%%%%%%%%

In summary, the availability of determinantal transition rate formulas opens up
a whole new area of applications of the Bethe ansatz with enormous potential for
important new results, among them results of relevance for experiments (neutron
scattering, NMR) on quasi-one-dimensional magnetic insulators.

%%%%%%%%%%%%%%%%%%%%%%%%%%%%%%%%%%%%%%%%%%%%%%%
%
\ack
%
%%%%%%%%%%%%%%%%%%%%%%%%%%%%%%%%%%%%%%%%%%%%%%%
Financial support from the DFG Schwerpunkt \textit{Kollektive
  Quantenzust{\"a}nde in elektronischen 1D {\"U}bergangsmetallverbindungen} (for
M.K.) is gratefully acknowledged. M.K. thanks Prof. Dr. Frommer and Prof. Dr.
Fabricius for useful discussions. D.B. thanks Dr. Fledderjohann for useful
discussions.
%%%%%%%%%%%%%%%%%%%%%%%%%%%%%%%%%%%%%%%%%%%%%
%\begin{appendix}
%%%%%%%%%%%%%%%%%%%%%%%%%%%%%%%%%%%%%%%%%%%%%
%
%\section{}\label{appA}
%
%%%%%%%%%%%%%%%%%%%%%%%%%%%%%%%%%%%%%%%%%%%%%
%
%\end{appendix}
%%%%%%%%%%%%%%%%%%%%%%%%%%%%%%%%%%%%%%%%%%%%%%%
%%%%%%%%%%%%%%%%%%%%%%%%%%%%%%%%%%%%%%%%%%%%%%%
\section*{References}
%\ifthenelse{\equal{\writer}{gerhard}}%
%{\bibliography{../references,notes}}%
%{\bibliography{/home/karbach/REFERENCES/references,notes}}

\begin{thebibliography}{100}
\bibitem{note4}
The most challenging part handling complex solutions is finding the complex
  roots of the Bethe ansatz equations.
\end{thebibliography}
%\bibliographystyle{jpa}

%%%%%%%%%%%%%%%%%%%%%%%%%%%%%%%%%%%%%%%%%%%%%%%
\end{document}